\documentclass[aps,prl,twocolumn,superscriptaddress,showkeys,amsfonts,amssymb,amsmath]{revtex4-1}

\usepackage[dvipdfmx]{graphicx}
\usepackage{color,amsmath,latexsym,amsfonts,amssymb}
\usepackage{inputenc}
\usepackage[colorlinks=true]{hyperref}
\usepackage{setspace}
\usepackage{multirow}
\usepackage{array}
\usepackage{tabularx}
\usepackage{bm}
\usepackage{color}

\newcommand {\bacovo} {BaCo$_2$V$_2$O$_8$}

\begin{document}

\title{Tomonaga-Luttinger liquid spin dynamics\\ in the quasi-one dimensional Ising-like antiferromagnet \bacovo}

\author{Quentin Faure}
\affiliation{Univ. Grenoble Alpes, CEA, INAC--MEM, Grenoble, France}
\affiliation{Univ. Grenoble Alpes, Inst NEEL, Grenoble, France}
\author{Shintaro Takayoshi}
\email[Corresponding author: ]{shintaro@pks.mpg.de}
\affiliation{Max Planck Institute for the Physics of Complex Systems, Dresden, Germany}
\affiliation{Department of Quantum Matter Physics, University of Geneva, Geneva, Switzerland}
\author{Virginie Simonet}
\affiliation{Univ. Grenoble Alpes, Inst NEEL, Grenoble, France}
\author{B\'{e}atrice Grenier}
\email[Corresponding author: ]{grenier@ill.fr}
\affiliation{Univ. Grenoble Alpes, CEA, INAC--MEM, Grenoble, France}
\author{Martin M{\aa}nsson}
\affiliation{Laboratory for Neutron Scattering and Imaging, Paul Scherrer Institute, Villigen PSI, Switzerland}
\affiliation{Department of Applied Physics, KTH Royal Institute of Technology, Kista, Stockholm, Sweden}
\author{Jonathan S.~White}
\affiliation{Laboratory for Neutron Scattering and Imaging, Paul Scherrer Institute, Villigen PSI, Switzerland}
\author{Gregory S.~Tucker}
\affiliation{Laboratory for Neutron Scattering and Imaging, Paul Scherrer Institute, Villigen PSI, Switzerland}
\affiliation{Laboratory for Quantum Magnetism, Institute of Physics, Ecole Polytechnique F\'ed\'erale de Lausanne (EPFL), Lausanne, Switzerland}
\author{Christian R\"uegg}
\affiliation{Laboratory for Neutron Scattering and Imaging, Paul Scherrer Institute, Villigen PSI, Switzerland}
\affiliation{Department of Quantum Matter Physics, University of Geneva, Geneva, Switzerland}
\affiliation{Neutrons and Muons Research Division, Paul Scherrer Institute, Villigen PSI, Switzerland}
\author{Pascal Lejay}
\affiliation{Univ. Grenoble Alpes, Inst NEEL, Grenoble, France}
\author{Thierry Giamarchi}
\affiliation{Department of Quantum Matter Physics, University of Geneva, Geneva, Switzerland}
\author{Sylvain Petit}
\affiliation{Laboratoire L\'eon Brillouin, CEA, CNRS, Universit\'e Paris-Saclay, CEA-Saclay, Gif-sur-Yvette, France}


\begin{abstract}
Combining inelastic neutron scattering and numerical simulations, we study the quasi-one dimensional Ising anisotropic quantum antiferromagnet \bacovo\ in a longitudinal magnetic field. This material shows a quantum phase transition from a N\'eel ordered phase at zero field to a longitudinal incommensurate spin density wave at a critical magnetic field of 3.8 T. Concomitantly the excitation gap almost closes and a fundamental reconfiguration of the spin dynamics occurs. These experimental results are well described by the universal Tomonaga-Luttinger liquid theory developed for interacting spinless fermions in one dimension. We especially observe the rise of mainly longitudinal excitations, a hallmark of the unconventional low-field regime in Ising-like quantum antiferromagnet chains.
\end{abstract}

\maketitle

Quantum magnets offer an extremely rich variety of phases ranging from the conventional long-range ordered ones, dubbed spin ``solids'', to various kinds of spin ``liquids''. In the latter, the excitations have often an unconventional nature such as a topological character or fractional quantum numbers. Among such systems, one dimensional (1D) quantum magnets are especially interesting in that the topological excitations are the norm rather than the exception, and because the interplay between exchange coupling and extremely strong quantum fluctuations due to the reduced dimensionality gives rise to profuse physical phenomena~\cite{giamarchi2004}. 

\begin{figure}
\centering
\includegraphics[width=8cm]{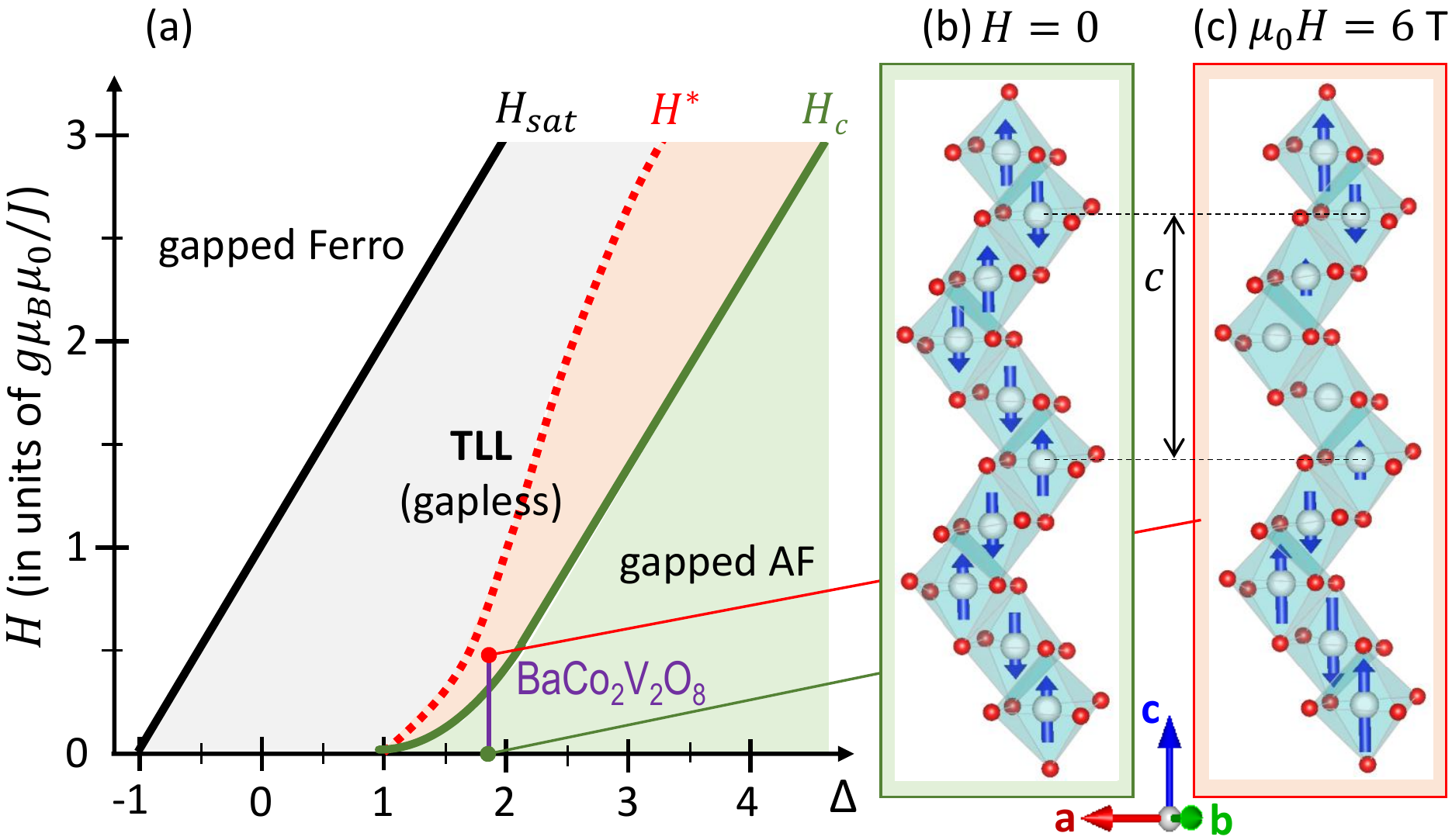}
\caption{(a) The ground state phase diagram of the spin-1/2 $XXZ$ chain under a longitudinal field with Hamiltonian~\eqref{eqXXZ}. The Heisenberg case corresponds to $\Delta=1$ and \bacovo\ to $\Delta=1.9$. The grey-shaded area ($H^*<H<H_{sat}$) is dominated by transverse spin-spin correlations and the red-shaded area ($H_c<H<H^*$) by longitudinal correlations. (b)-(c) Magnetic structure (blue arrows) of a single Co$^{2+}$ screw chain of \bacovo\ (blue and red spheres are Co and O respectively) at (b) $H=0$ in the gapped N\'eel phase ($H<H_c$) and (c) $\mu_0H=6$~T in the low field regime of the TLL phase ($H_c<H<H^*$). The amplitude of the magnetic moments in (c) is multiplied by 5 for clarity.}
\label{fig.1}
\end{figure}

On the experimental front, the recent realization of quantum magnets with relatively weak magnetic exchange has paved a new avenue to an efficient manipulation of systems with realizable magnetic fields, enabling novel phases and phenomena to be probed experimentally. Plentiful examples of such successful investigations exist, e.g. scaling properties of Bose-Einstein condensation~\cite{giamarchi-ruegg-magnonBEC2008,batista-RMP-magnonBEC2014}, quantitative tests of Tomonaga-Luttinger liquid (TLL) theory~\cite{klanjsek-PRL-LLinladder2008,bouillot-PRB-ladder2011,schmidiger-PRL-DIMPY2013}, scaling properties at quantum critical points~\cite{zheludev-PRB-CriticalScale2017,zheludev-PRL-CriticalBPCB2018}, fractionalized excitations~\cite{broholm-takagi-PRL-SrCuO2-spinon2004,thielemann-PRL-fractional-ladder2009}, topological phase transitions~\cite{faure2018}, other exotic excitations~\cite{zheludev-giamarchi-tsvelik-PRB-ladder2013,grenier2015,bera2017,wang2018}. The effect of an external magnetic field competing with the excitation gap associated to rung-singlets~\cite{klanjsek-PRL-LLinladder2008} or to the Haldane state~\cite{renard-PRL-MagneticFieldHaldaneChain1989} for instance is especially interesting. Quite remarkably, all these transitions fall into the same universality class, the so called Pokrovsky-Talapov commensurate-incommensurate (C-IC) phase transition~\cite{Talapov1979,giamarchi2004}, which is also pertinent to the Mott transition in itinerant electronic systems. Hence there is a considerable interest in experimental analyses of such phenomena, and investigations have been conducted in systems such as bosons in a periodic lattice~\cite{naegerl-Nature-pinning-transition2010,modugno-PRA-BosonMott2016}, spin-1 chains~\cite{zvyagin-PRL-DTN-magfield2007} and spin-1/2 ladders~\cite{klanjsek-PRL-LLinladder2008,bouillot-PRB-ladder2011}. However, in these realizations, magnetic excitations in the IC phase are dominated by spin-spin correlations transverse to the applied field, and a study of the opposite and more exotic case, where the longitudinal excitations are dominant, is still lacking. 

In this paper, we focus on this particular case. We investigate the Ising-like compound \bacovo\ under a magnetic field along the anisotropy axis by combining inelastic neutron scattering experiments and numerical simulations. We show that the quantum phase transition provoked by a longitudinal field of 3.8 T is indeed in the C-IC universality class through the analysis of spin-spin dynamical correlations. Furthermore, we demonstrate that most of the spectral weight in the IC phase consists in {\it longitudinal} excitations, which are a strong fingerprint of TLL dynamics with IC solitonic excitations.

\bacovo\ consists of screw chains of Co$^{2+}$ ions running along the fourfold $c$-axis of a body-centered tetragonal structure [Fig.~\ref{fig.1}(b)] \cite{wichmann1986}. Due to an anisotropic $g$ tensor~\cite{kimura2006}, the Co$^{2+}$ magnetic moments are described effectively by weakly coupled spin-1/2 $XXZ$ (Ising-like) chains~\cite{abragam1951}. The Hamiltonian includes intrachain and interchain interactions ${\cal H}=\sum_{\mu}{\cal H}_{{\rm intra},\mu}+{\cal H}_{\rm inter}$, which write 
\begin{align}
{\cal H}_{{\rm intra},\mu} =& J \sum_{n} ( S_{n,\mu}^x S_{n+1,\mu}^x 
+ S_{n,\mu}^y S_{n+1,\mu}^y+\Delta S_{n,\mu}^z S_{n+1,\mu}^z) \nonumber\\
&- g_{zz}\mu_B \mu_0H\sum_{n} S^{z}_{n,\mu},
\label{eqXXZ}
\end{align}
and ${\cal H}_{\rm inter}=J'\sum_{n}\sum_{\langle\mu,\nu\rangle}{\bf S}_{n,\mu}\cdot{\bf S}_{n,\nu}$. Here ${\bf S}_{n,\mu}$ is a spin-1/2 operator, $n$ the site index along the chain, $\mu,\nu$ label different chains, $J(>0)$ is the antiferromagnetic (AF) intrachain interaction, and $\Delta$ the Ising anisotropy. $g_{zz}\mu_B \mu_0H\sum_{n} S^{z}_{n,\mu}$ is the Zeeman term from the longitudinal field, with $g_{zz}$ the Land\'e factor and $\mu_{B}$ the Bohr magneton. The $a,b,c$ crystallographic axes coincide with the spin $x,y,z$ axes, respectively. The interchain coupling is treated by mean field theory~\cite{supmat1}. At $H=0$ and $T \leq T_N$  ($T_N=5.4$~K), \bacovo\ is in a gapped AF phase and the magnetic moments point along the Ising $c$-axis [Fig.~\ref{fig.1}(b)]. The elementary excitations are spinons, which are confined by the interchain coupling to form spinon bound states. They give rise to two series of discretized energy levels dispersing along the $c$-axis (and only weakly in the perpendicular directions), which have longitudinal ($\Delta S^z=0$) and transverse ($\Delta S^z=\pm1$) character with respect to the anisotropy axis~\cite{grenier2015}. The ground state phase diagram of a single spin-1/2 $XXZ$ chain under the application of a longitudinal magnetic field is shown in Fig.~\ref{fig.1}(a). In the Ising-like case ($\Delta>1$), $H>H_c$ is required to enter the TLL phase and close the excitation gap. The TLL phase is characterized by spatial spin-spin correlations transverse $C^{xx}(r)\equiv\langle S^x_rS^x_0\rangle\propto r^{-1/(2K)}$ and longitudinal $C^{zz}(r)\equiv\langle S^z_rS^z_0\rangle-m_z^2\propto r^{-2K}$ to the field direction, where $m_z$ is the field-induced uniform magnetization per site. The decay of $C^{zz}(r)$ and $C^{xx}(r)$ are dictated by the TLL parameter $K$. The field dependence of $K$ causes a crossover at $H^*$ from a low-field regime [red-shaded area in Fig.~\ref{fig.1}(a)] where the physics is dominated by $C^{zz}(r)$ to a high-field regime [grey-shaded area in Fig.~\ref{fig.1}(a)] dominated by $C^{xx}(r)$. The dispersion of low-energy excitations is expected to become gapless at both C and IC wave vectors $q=\pi,2\pi m_z$ for transverse excitations (captured by the space-time correlation $\langle S^x_r(t)S^x_0(0)\rangle$), and at $q=0,\pi(1-2m_z)$ for longitudinal excitations (captured by $\langle S^z_r(t)S^z_0(0)\rangle$)~\cite{muller1981,chitra1995,chitra1997,fath2003}. For \bacovo\ ($\Delta=1.9$), the quantum phase transition occurs at $\mu_0H_{c}=3.8$~T from the N\'eel phase to the longitudinal spin density wave (LSDW) with an IC wave vector, both ordered phases stabilized by weak interchain couplings. In the latter phase, the magnetic moments are parallel to the field (and Ising) direction while their amplitude is spatially modulated [Fig.~\ref{fig.1}(c)] ~\cite{kimura2008a,kimura2008b,canevet2013,supmat1}. When the external field is further increased, the LSDW phase is replaced by a canted AF order with staggered moments perpendicular to the $c$-axis above $\mu_0H^*\approx9$~T~\cite{grenierPRB2015,klanjsek2015}, which corresponds to the crossover from the TLL longitudinal to transverse-dominant correlations, before the magnetization saturates at higher field ($H_{sat}$).

\begin{figure}
\centering
\includegraphics[width=8cm]{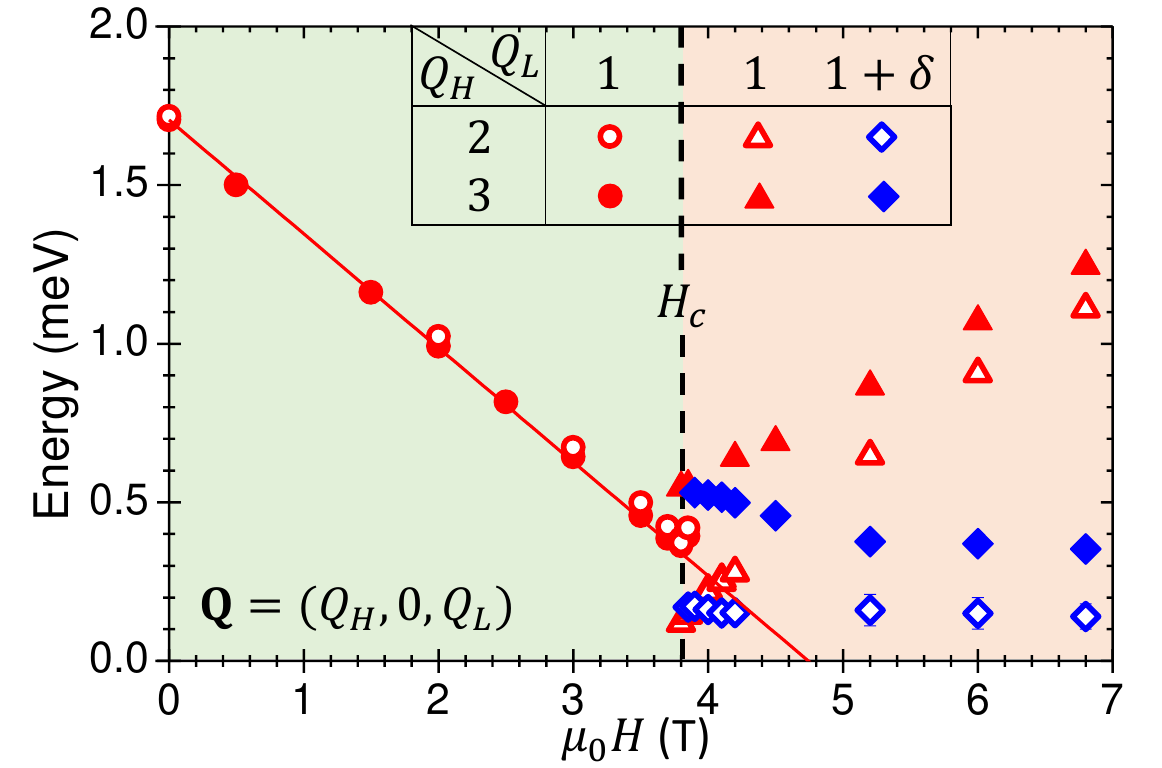}
\caption{Field dependence of low energy magnetic excitations in \bacovo\, across the quantum phase transition occurring for ${\bf H}\parallel {\bf c}$. Open and closed symbols correspond respectively to magnetic and crystallographic positions. Circles correspond to AF ${\bf Q} = (2,0,1)$ and ZC ${\bf Q} = (3,0,1)$ positions. Triangles denote the same AF and ZC positions in the IC phase. Diamonds correspond to the associated satellites ${\bf Q} = (2,0,1+\delta)$ and ZC  IC ${\bf Q} = (3,0,1+\delta)$. The critical field is indicated by  the dashed black line.}
\label{fig.2}
\end{figure}

To probe the transition from the N\'eel to LSDW phase and their spin dynamics, we performed inelastic neutron scattering experiments at the cold-neutron triple axis spectrometer TASP (PSI, Switzerland). We used a horizontal cryomagnet, applying magnetic fields up to 6.8 T. Two \bacovo\ single crystals, grown by floating zone, were co-aligned with an accuracy better than $1^{\circ}$. The magnetic field was applied along the $c^*-$axis of the $(a^*, c^*)$ scattering plane, hence along the magnetic moment direction. The data were measured at the base temperature of 150~mK with various fixed final wave vectors ranging from 1.06 to 1.3~\AA$^{-1}$ (yielding an energy resolution from 70 to 150 $\mu$eV). In \bacovo, the crystallographic zone centers (ZC) are at ${\bf Q} = (h,k,l)$ positions with $h+k+l={\rm even}$. The magnetic Bragg peaks of the N\'eel phase appear at the AF points ${\bf Q} = (h+1,k,l)$ corresponding to the ${\bf k}_{AF}=(1, 0, 0)$ propagation vector \cite{canevet2013}. The presence of four screw-chains per unit cell folds the excitation branches and replication from the ZC positions is added to the usual contribution from AF points. 

Energy scans with constant $Q$ have first been recorded for different magnetic fields at the AF position ${\bf Q} =(2, 0, 1)$. At $H=0$, the measured lowest energy peak corresponds to the doubly degenerate transverse excitation~\cite{grenier2015}. The field produces a Zeeman splitting that lifts this degeneracy \cite{kimura2007,faure2018}, leading to the linear decrease of the lowest transverse mode up to the transition at $H_c$, as observed in Fig.~\ref{fig.2} (red open circles). The same feature is seen at ZC wave vectors (red closed circles) due to the folding.

\begin{figure}
\centering
\includegraphics[width=\linewidth]{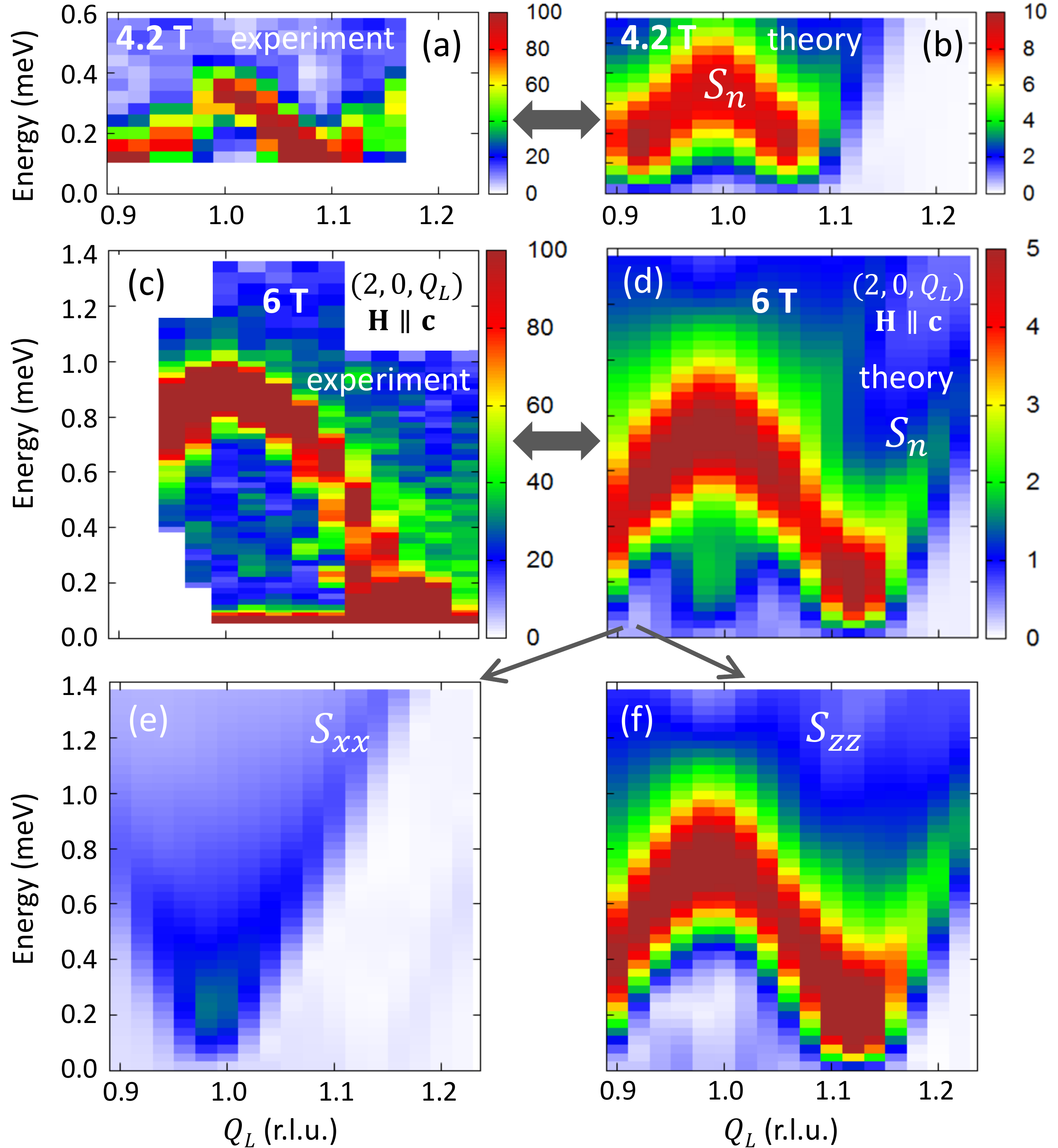}
\caption{Inelastic scattering intensity maps showing the intrachain dispersion of the magnetic excitations around the AF point ${\bf Q}=(2,0,1)$ in a longitudinal field of (a) 4.2~T and (c) 6~T, obtained experimentally from a series of constant-$Q_L$ energy scans. They are compared with numerically calculated scattering cross sections $S_n$~\cite{supmat1} at (b) 4.2~T and (d) 6~T, which are the superposition of (e) transverse $S_{xx}$ and (f) longitudinal $S_{zz}$ dynamical structure factors.}
\label{fig.3}
\end{figure}

In the LSDW phase, the propagation vector becomes ${\bf k}_{LSDW}=(1, 0, \delta)$. The field dependence of the IC modulation $\delta$ has been determined from $Q_L$-scans. In agreement with the TLL theory and previous report~\cite{canevet2013}, we have found that it increases with the field as $\delta=2\pi m_z$, i.e., the period for the spatial modulation of the magnetic moments becomes shorter~ \cite{supmat1}. The transition at $H_c$ into the LSDW phase also manifests as a change of magnetic excitation spectrum [from circles to triangles in Fig.~\ref{fig.2}]. To obtain the overall behavior of spin dynamics in this LSDW phase, constant-$Q_L$ energy scans have been collected along the $c^*$ direction across the AF point ${\bf Q}=(2,0,1)$ at $\mu_0H=4.2$ and $6$~T. Figures~\ref{fig.3}(a) and \ref{fig.3}(c) show the corresponding maps as a function of energy transfer and $Q_L$. At 4.2~T, a strong excitation is observed, forming an arch bridging the IC positions $(2,0,1\pm \delta)$ over the AF center $(2,0,1)$. The dispersion has minima at the IC positions of the LSDW phase, which is a key signature of this field-induced TLL phase. The data show that the arch-like dispersion expands from 4.2~T to 6~T, while $\delta$ becomes about twice larger: The energy minimum at $(2,0,1\pm \delta)$ remains equal to $\approx$0.1~meV while the energy at the AF point increases.

\begin{figure*}[t]
\centering
\includegraphics[width=\linewidth]{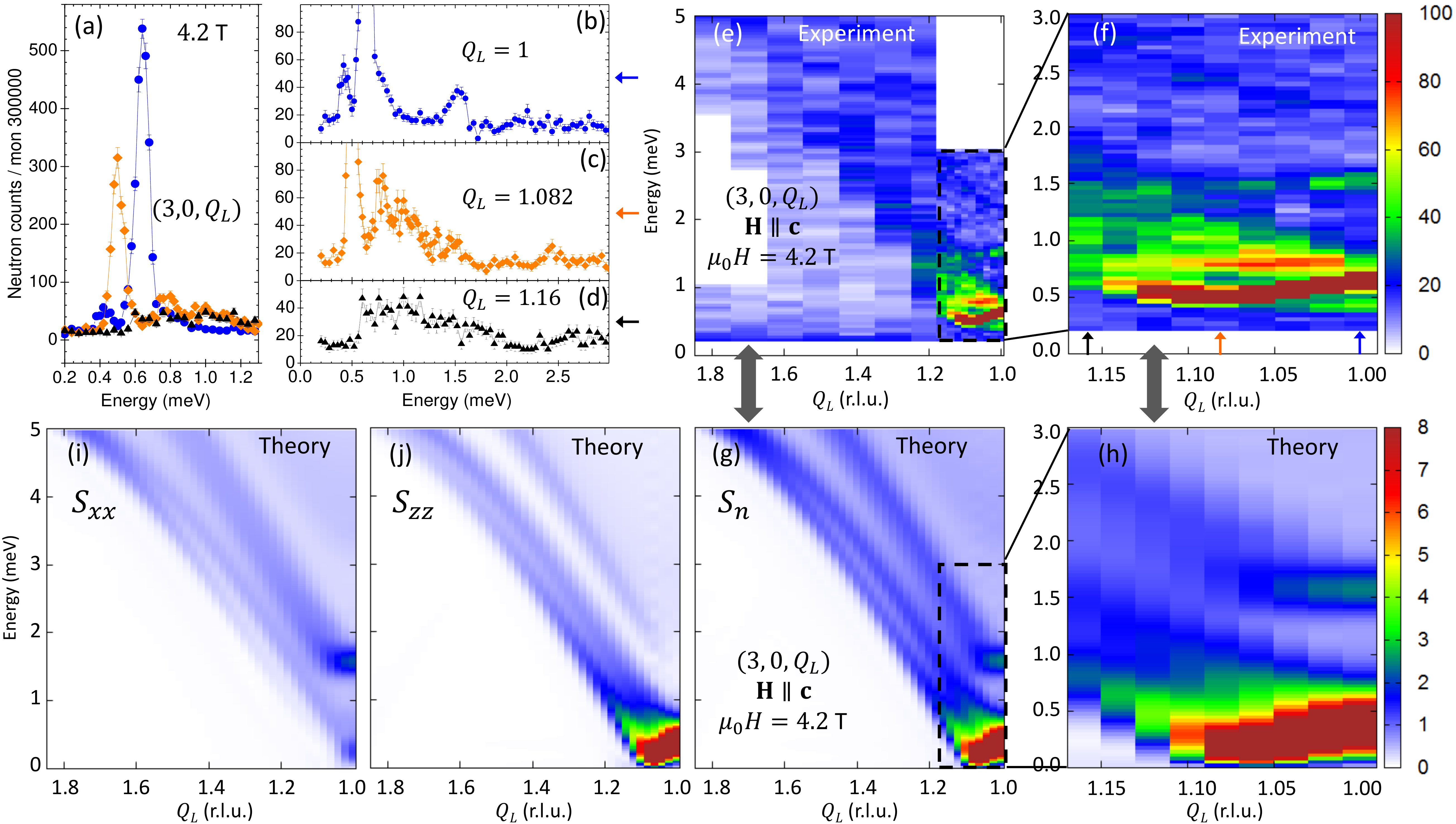}
\caption{Spin dynamics in the LSDW phase of \bacovo\ in the vicinity of a ZC position at $\mu_0H = 4.2$~T and $T = 150$~mK. Energy scans at three different scattering vectors (b) the ZC position $(3,0,1)$, (c) its satellite $(3,0,1+\delta)$ with $\delta = 0.082$, and (d) further away along $c^*$. (a) Low energy part of these scans. (e) Inelastic scattering intensity map obtained from constant-$Q_L$ energy scans as (a)-(d), which shows the dispersion of the excitations along the $c^\star$ direction, and is zoomed in (f). (g)-(h) Numerically calculated intensity color maps to be compared with experimental maps (e)-(f). In both the experimental and theoretical maps, the color scale is saturated in order to emphasize the weaker modes. (i) Transverse $S_{xx}$ and (j) longitudinal $S_{zz}$ components of the numerically calculated intensity color map (g).}
\label{fig.4}
\end{figure*}

From the above-mentioned folding, replications are observed around the ZC position ${\bf Q} = (3, 0, 1)$. This is illustrated in Figs.~\ref{fig.4}(a)-\ref{fig.4}(d) showing individual constant-$Q_L$ energy scans through ${\bf Q}=(3,0,Q_L)$ positions with $Q_L$ ranging between $1$ and $1.16$ and energy up to 5~meV at 4.2~T. The map gathering such scans is displayed in Fig.~\ref{fig.4}(e) with a zoom in Fig.~\ref{fig.4}(f). These results show that most of the intensity is  concentrated in an arch-like excitation with minimum energy of the dispersion $\simeq 0.65$~meV at $(3,0,1+\delta)$, the satellite position of the LSDW phase. At 6~T, the intense arch feature expands similarly to the result around the AF position~\cite{supmat1}. Weaker excitations are also visible around 0.4, 0.8, and 1.5~meV. Further away from the ZC position along $Q_L$, only a broad feature remains, possibly corresponding to a continuum of excitations [see Fig.~\ref{fig.4}(d) for $Q_L=1.16\simeq 1+2 \delta$]. Although the excitations in the AF and ZC regions show strong similarities, the energy gap at the IC wave vector is significantly smaller at the AF satellite than at the ZC one. This is also visible in Fig.~\ref{fig.2} displaying the energy of the intense modes at the two IC positions $(2, 0, 1+\delta)$ and $(3, 0, 1+\delta)$. This is ascribed to the finite dispersion perpendicular to the chain direction caused by the interchain coupling, and also observed in zero field~\cite{grenier2015}.

Aiming at a deeper understanding of the spin dynamics in the LSDW phase, we performed numerical simulations of the $XXZ$ model with a longitudinal magnetic field [Eq.~\eqref{eqXXZ}]. We obtained the ground state of the system by density matrix renormalization group~\cite{white1992} and calculated the retarded correlation function by time-evolving block decimation~\cite{vidal2003}. The inelastic neutron scattering cross section $S_n$ was derived as the Fourier transform of this correlation function \cite{faure2018,takayoshi2018}. The calculations were performed by considering the full magnetic structure factor of \bacovo\ with the values $J=3.05$~meV and $\Delta=1.9$~\cite{J} obtained from our previous investigation~\cite{faure2018}. The agreement is best for interchain coupling $J'=0$ and deteriorates with increasing it, especially near the C-IC transition point~\cite{supmat1}, in contract with our previous estimation of $J'=0.17$~meV~\cite{grenier2015}. This may be due to a mean-field overestimation of its effect particularly in the critical region or to its possible dependence on the longitudinal field since it is an effective coupling derived from a complex set of interactions~\cite{klanjsek2015}. All the numerical calculations presented here were therefore performed with $J'=0$.

The calculated field dependence of $\delta(H)$ globally agrees with the experiment except near the transition~\cite{supmat1}. We present the comparison of the measured vs calculated excitation spectra in Figs.~\ref{fig.3}(a) vs \ref{fig.3}(b) and \ref{fig.3}(c) vs \ref{fig.3}(d) around the AF position at 4.2 and 6~T respectively, as well as in Figs.~\ref{fig.4}(e)-\ref{fig.4}(f) vs \ref{fig.4}(g)-\ref{fig.4}(h) around the ZC position at 4.2~T.  Note that the calculated peaks are broadened (0.3~meV resolution) compared to the experimental ones due to the finite time effect, i.e. the limitations of the calculations within the finite real time domain $0\leq t\leq T$. The main features, i.e. the dispersion of the low energy excitation bridging the two neighboring IC wave vectors and its spectral weight, are well reproduced. The relative intensity of the weaker branches at 0.8 and 1.5~meV at $(3, 0, 1+\delta)$ is less accurately reproduced, maybe due to the omission of interchain interaction in the calculations. 

The nature of the fluctuations can be further analyzed by the numerically calculated transverse and longitudinal parts of the dynamical structure factor, $S_{xx}$ and $S_{zz}$, which are shown in Figs.~\ref{fig.3}(e)-\ref{fig.3}(f) around $(2,0,Q_L)$ at 6~T and in Figs.~\ref{fig.4}(i)-\ref{fig.4}(j) around $(3,0,Q_L)$ at 4.2~T. The most striking result is that the arch-like excitation has longitudinal character around both AF and ZC positions. For $(3,0,Q_L)$ at 4.2~T, the weaker transverse excitations $S_{xx}$ give rise to two branches going softer toward the C position with minimum energies close to zero and 1.5~meV [Fig.~\ref{fig.4}(i)], both of which are seen in the experimental data of Fig.~\ref{fig.4}(f). This result proves that the spin dynamics is dominated by longitudinal fluctuations strongly excited at the IC wave vectors near the AF positions, which replicate around the ZC ones.

A recent THz spectroscopy investigation of the spin dynamics was performed under a longitudinal magnetic field in the gapless regime of SrCo$_2$V$_2$O$_8$, the sister compound of \bacovo~\cite{yang2017,wang2018}. In this experiment, only transverse excitations ($S_{xx}$) at C positions could be probed, such as string and (anti)psinon-(anti)psinon, dressing the field-polarized ground state of 1D quantum antiferromagnets described by the Bethe Ansatz~\cite{karbach2002,kohno2009}. Our neutron spectroscopy study opens up new avenues. We could first follow the dispersion in reciprocal space of the  psinon-psinon and 2-string excitations corresponding to the weak transverse modes visible near zero and at 1.5 meV for $Q_L= 1$ in Fig.~\ref{fig.4}(f). Moreover, both transverse and longitudinal fluctuations could be probed and we have proven that most of the intensity actually comes from longitudinal excitations missed by THz spectroscopy. This finding is essential to understand a growing number of experiments performed on similar systems with other probes. Our results finally pave the way to further investigations of unexplored regimes of the TLL physics in spin systems, such as the crossover from  longitudinal to transverse dominant spin-spin correlations at higher magnetic field or the influence of interchain interactions.

In summary, our combined neutron scattering and numerical investigations of the LSDW phase in \bacovo\ show that the quantum phase transition from the N\'eel to LSDW phase is described by the $XXZ$ model. Clear Tomonaga-Luttinger liquid signatures are observed such as the field-dependent incommensurability of the low energy excitations and the arch-like dispersion. The most striking result concerns the longitudinal nature of the excitations in the LSDW phase, which is a remarkable quantum signature of the field-induced TLL in Ising-like spin 1/2 1D antiferromagnets. 

\acknowledgments


\clearpage
\widetext
\begin{center}
\textbf{\large Supplementary Information: \\Topological quantum phase transition \\ in the Ising-like antiferromagnetic spin chain BaCo$_2$V$_2$O$_8$}
\end{center}

\setcounter{equation}{0}
\setcounter{figure}{0}
\setcounter{table}{0}
\setcounter{page}{1}
\makeatletter
\renewcommand{\theequation}{S\arabic{equation}}
\renewcommand{\thefigure}{S\arabic{figure}}
\renewcommand{\bibnumfmt}[1]{[S#1]}
\renewcommand{\citenumfont}[1]{S#1}



\section{Field-dependence of the incommensurability}

\begin{figure}[htb]
\centering
\includegraphics[width=10.cm]{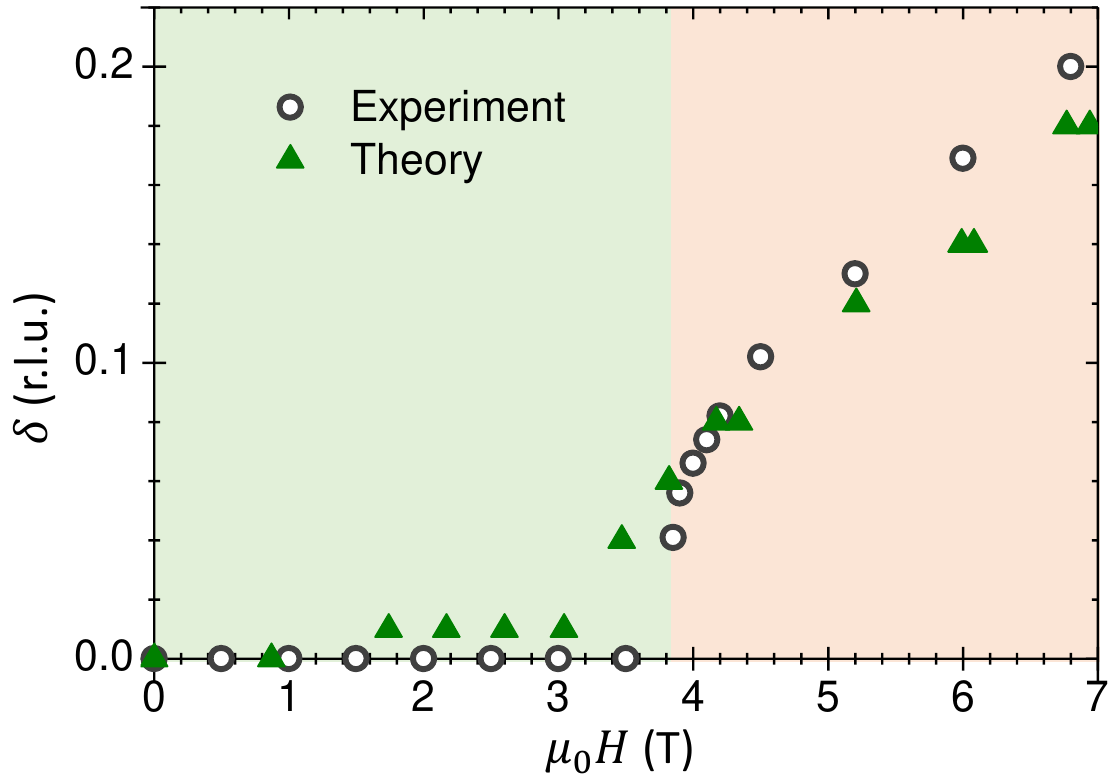}
\caption{Incommensurate modulation $\delta$ of the LSDW phase, characterized by the propagation vector ${\bf k}_{LSDW}=(1, 0, \delta)$, as a function of the longitudinal magnetic field. Black circles show the experimental data and green triangles show the DMRG calculations. The red and green background colors correspond to those used in Figs.~1 and 2 of the main paper.}
\label{fig1SM}
\end{figure}

Figure~\ref{fig1SM} shows the measured and calculated field-dependence of the incommensurability wave vector in the longitudinal spin density wave (LSDW) phase. We have calculated the ground state numerically by density matrix renormalization group (DMRG) for a finite size system consisting in 200 sites and obtained the local magnetization $M_{r}^{z}\equiv\langle S_{r}^{z}\rangle$. $\delta$ is determined from the peak that occurs in its Fourier transform $M^{z}(q)\equiv|\sum_{r}e^{iqr}M_{r}^{z}|$.

\section{Nature of the LSDW phase}

A simple interpretation for the nature of the LSDW phase can be obtained by considering the softening of the lowest mode with increasing the external magnetic field. At zero magnetic field, the lowest energy excitation is the doubly degenerate transverse mode ($\Delta S^z=\pm 1$), which splits due to the Zeeman effect. The energy of an excitation with $\Delta S^z=+1$ decreases while that with $\Delta S^z=-1$ increases. When the excitation gap is closed, the domain wall excitations with $\Delta S^z=+1$ condensate and the quantum phase transition happens. The number of these domain walls proliferates with increasing the magnetic field so as to minimize the Zeeman energy. This increase of the number of domain walls is related with the decrease of their average distance of separation $1/\delta$. The domain walls actually have some intrinsic width as shown by nuclear magnetic resonance (NMR) measurements~\cite{MK}, so that this array of walls coincides with the spin density wave deduced from neutron diffraction. W
e looked by neutron diffraction measurements for third order harmonics in the LSDW phase. Their presence would demonstrate a squaring of the sinusoidal amplitude modulation of the magnetic structure. However, such Bragg reflections could not be observed. The analysis of the error bars then shows that they must be at least 30 times smaller than the first order harmonics. This remains consistent with the expectations from the NMR line profile.


\section{Measured spin-dynamics at 6 T along (3, 0, $Q_L$)}

\begin{figure}[htb]
\centering
\includegraphics[width=10.cm]{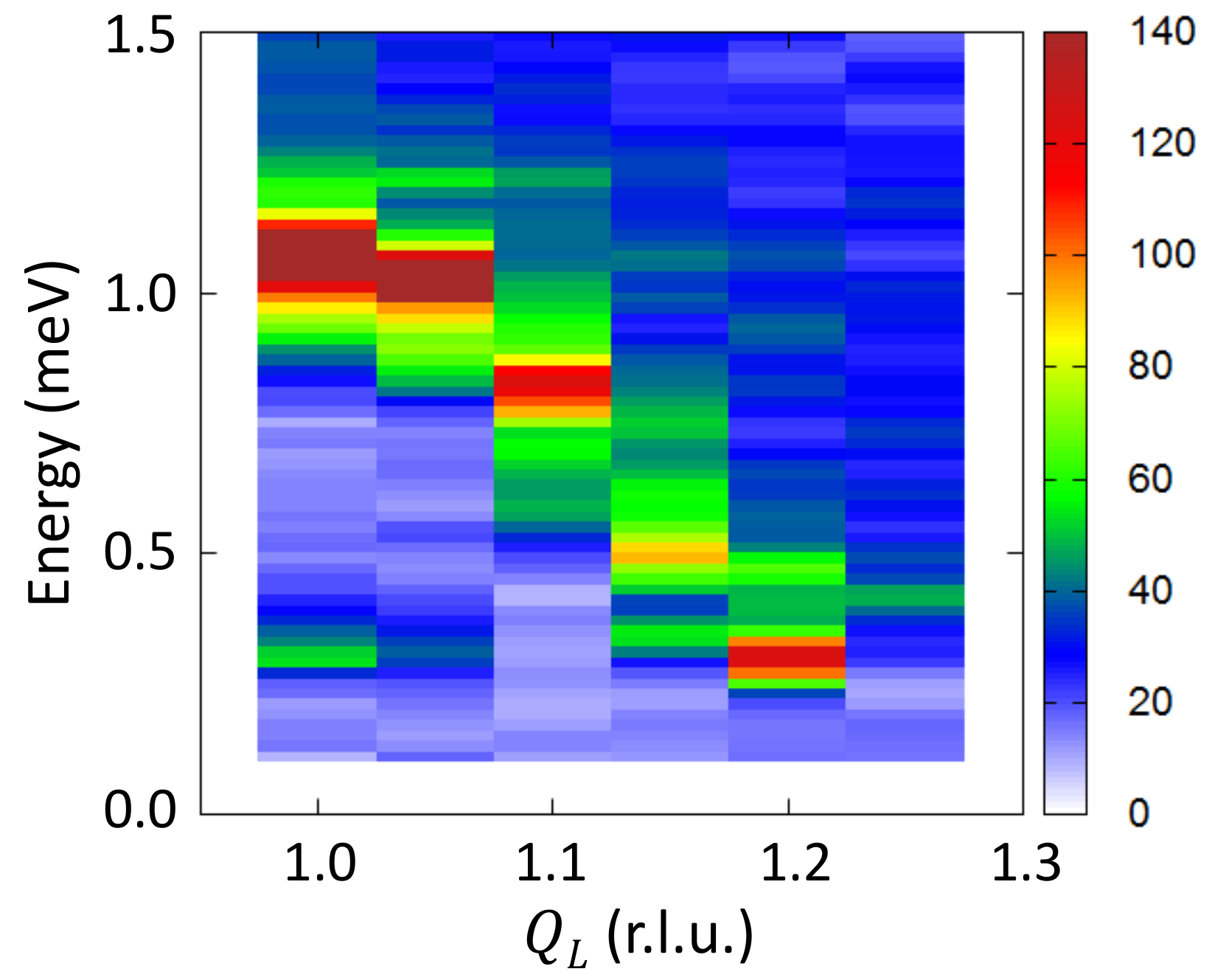}
\caption{Inelastic scattering intensity map showing the intrachain dispersion of the magnetic excitations along $\boldsymbol{Q}=(3,0,Q_L)$ around the ZC point $Q_L=1$ in a longitudinal field of 6~T. This experimental map was obtained from a series of $Q$ constant energy scans.}
\label{fig2SM}
\end{figure}

Constant-$Q$ energy scans have been recorded at $\mu_{0}H=6$ T along the $c^*$ direction across points $\boldsymbol{Q}=(3,0,Q_L)$ with $Q_L$ ranging between 1 [zone center (ZC) point] and 1.25. The resulting intensity map as a function of energy transfer and $Q_L$ is shown Fig.~\ref{fig2SM}. It features an intense arch-like excitation, which takes the minimum at the incommensurate positions near the ZC position $(3,0,1)$.

\section{Numerical simulations}

In this section, we explain the method used to perform the numerical simulations. The principle of the calculations are the same as in Refs.~\cite{faure2018,takayoshi2018}. Note that \bacovo\ consists of the stacking of Co chains, each of which can be considered as a spin-1/2 Heisenberg model with Ising (easy-axis) anisotropy. Taking into account the interchain coupling by the mean field theory, we obtain an effective one-dimensional Hamiltonian
\begin{equation}
 {\cal H}_{\rm eff} = J \sum_{n} ( S_{n}^x S_{n+1}^x
+ S_{n}^y S_{n+1}^y+\Delta S_{n}^z S_{n+1}^z) - g_{zz}\mu_{B}\mu_0 H\sum_{n} S^{z}_{n}
+ J'\sum_{n}\langle S^{z}_{n}\rangle S^{z}_{n}.
\label{eq:Hamil1Deff}
\end{equation}
Here, $x, y$ and $z$ coincide with the $a, b, c$ crystallographic axes. The local magnetization $\langle S^{z}_{n}\rangle$ is determined self-consistently. The parameters $J=3.05$ meV, $\Delta=1.9$ and $g$ factor along the $z$ axis $g_{zz}=6.07$ were determined so that they reproduce the neutron cross-section in zero-field at the scattering vector $\boldsymbol{Q}=(2, 0, 0)$ in the unit of $(2\pi/a,2\pi/b,2\pi/c)$, where $a,b,c$ are the lattice constants~\cite{faure2018}. Note that a different convention was used in Ref.~\cite{faure2018}, explaining the different numerical values of $J$: in the present paper, $J$ replaces $\epsilon J$ and $J\Delta$ replaces $J$ with $\Delta = 1/\epsilon = 1.9$. The differential neutron scattering cross section is represented as
\begin{equation}
 S_{n}(\boldsymbol{Q},\omega)\equiv\frac{d^{2}\sigma}{d\Omega dE}\propto
   \frac{|\boldsymbol{q}'|}{|\boldsymbol{q}|}\sum_{\alpha,\beta=x,y,z}
   \Big(\delta_{\alpha\beta}-\frac{Q_{\alpha}Q_{\beta}}{|\boldsymbol{Q}|^{2}}\Big)
   |F(\textbf{Q})|^{2}S_{\alpha\beta}(\boldsymbol{Q},\omega),
\label{eq:CrossSec}
\end{equation}
where $F(\boldsymbol{Q})$ is the magnetic form factor and $\boldsymbol{q},\boldsymbol{q}'$ are the initial and final wave vectors, respectively ($\boldsymbol{Q}=\boldsymbol{q}-\boldsymbol{q}'$). The dynamical structure factor $S_{\alpha\beta}(\boldsymbol{Q},\omega)$ is given as
\begin{equation}
 S_{\alpha\beta}(\boldsymbol{Q},\omega)
   =\bigg|\mathrm{Im}\int dt\sum_{\boldsymbol{r}}
   e^{i(\omega t-\boldsymbol{Q}\cdot\boldsymbol{r})}
   C_{\rm ret}^{\alpha\beta}(\boldsymbol{r},t)\bigg|.
\label{eq:DSF}
\end{equation}
Here $C_{\rm ret}^{\alpha\beta}(\boldsymbol{r},t)$ is the retarded correlation function
\begin{equation}
 C_{\rm ret}^{\alpha\beta}(\boldsymbol{r},t)
   =-i\vartheta_{\mathrm{step}}(t)
     \langle[S^{\alpha}(\boldsymbol{r},t),S^{\beta}(\boldsymbol{0},0)]\rangle,
\nonumber
\end{equation}
where $\vartheta_{\mathrm{step}}(t)$ is the step function.
When the system has a rotational symmetry around the $z$ axis, as is the case of the $XXZ$ chain under a longitudinal field~\eqref{eq:Hamil1Deff}, Eq.~\eqref{eq:CrossSec} is recast into
\begin{equation}
 S_{n}(\boldsymbol{Q},\omega)\propto
   \frac{|\boldsymbol{q}'|}{|\boldsymbol{q}|}|F(\textbf{Q})|^{2}
   \Big[\Big(1+\frac{Q_{z}^{2}}{|\boldsymbol{Q}|^{2}}\Big)
   S_{xx}(\boldsymbol{Q},\omega)
   +\Big(1-\frac{Q_{z}^{2}}{|\boldsymbol{Q}|^{2}}\Big)
   S_{zz}(\boldsymbol{Q},\omega)\Big].
\nonumber
\end{equation}

We first obtain the ground state using DMRG~\cite{white1992}, then perform the time-evolution with time-dependent block decimation (TEBD)~\cite{vidal2003} and calculate space-time correlation functions for the Hamiltonian~\eqref{eq:Hamil1Deff}. In the calculations, the system size is $N=200$ and time interval is taken to be $0\leq t\leq 60J^{-1}$ with the discretization $dt=0.05J^{-1}$. The truncation dimension (i.e., the bond dimension of matrix product states) is $M=60$. For the Fourier transform in Eq.~\eqref{eq:DSF}, the summation is taken over the actual positions $\boldsymbol{r}$ of Co$^{2+}$ ions.

\section{Effects of the interchain interaction in the numerical calculations}

\begin{figure}[htb]
\centering
\includegraphics[width=18cm]{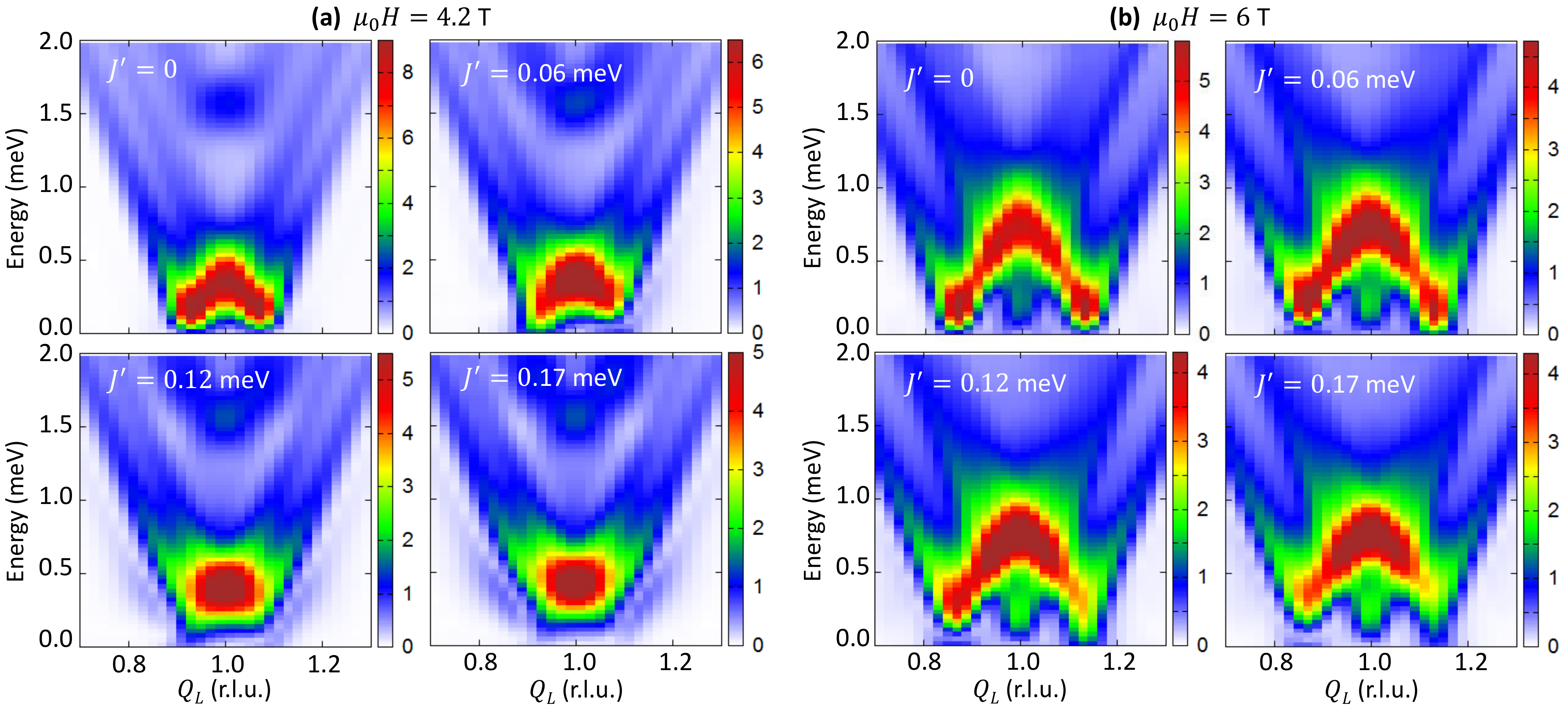}
\caption{The numerical results for the scattering cross section~\eqref{eq:CrossSec} along $(2,0,Q_L)$ around the AF position $Q_L=1$ in a longitudinal magnetic field of (a) 4.2~T and (b) 6 T for four values of the interchain interaction $J'$ increasing from 0 to 0.17 meV.}
\label{fig3SM}
\end{figure}

In this section, we examine the effects of the interchain interaction. Although it is likely that the interchain coupling consists of a complex set of interactions including further than the nearest neighbor~\cite{klanjsek2015}, we consider for simplicity the interchain coupling only between the nearest neighbor sites and treated it in a mean-field theory as stated in the previous section,
\begin{equation}
 {\cal H}_{\rm inter}=J'\sum_{n}\sum_{\langle\mu,\nu\rangle}
\boldsymbol{S}_{n,\mu}\cdot\boldsymbol{S}_{n,\nu}
\simeq J'\sum_{n,\mu}
\boldsymbol{S}_{n,\mu}\cdot\langle\boldsymbol{S}_{n}\rangle.
\end{equation}
In Fig.~\ref{fig3SM}, we show the results of numerical calculations around the AF position $\boldsymbol{Q}=(2,0,1)$ for $\mu_{0}H=4.2$~T [Fig.~\ref{fig3SM}(a)] and 6~T [Fig.~\ref{fig3SM}(b)] with varying $J'$ from 0 to 0.17 meV. The agreement between the numerics and the experimental data, presented in the main article and in Fig.~\ref{fig2SM} is best for $J'=0$ and worsens with increasing $J'$. The deviation becomes larger as $H$ approaches the critical field as far as the spectral weight distribution is concerned. With increasing $J'$, we can see that the N\'eel order (the weight at $Q_L=1$) is reinforced to the detriment of the incommensurate spin density wave. This is because an effective staggered field is induced by the N\'eel order through the interchain coupling. This results in an overestimation of the N\'eel order, and this effect becomes stronger as $H$ is closer to the phase transition point.

\end{document}